\documentclass[twocolumn,english,prl,superscriptaddress]{revtex4-2}
\usepackage[T1]{fontenc}
\usepackage[utf8]{inputenc}
\usepackage{amsmath,amssymb,graphicx,babel,braket,nicefrac}
\usepackage[svgnames]{xcolor}
\usepackage[unicode=true,pdfusetitle,
bookmarks=true,bookmarksnumbered=false,bookmarksopen=true,bookmarksopenlevel=1,breaklinks=false,pdfborder={0 0 0},pdfborderstyle={},colorlinks=true,linkcolor=DarkBlue,citecolor=DarkBlue,filecolor=DarkBlue,urlcolor=DarkBlue]{hyperref}
\usepackage{orcidlink}
\newcommand{\figref}[2]{\hyperref[#1]{\ref{#1}(#2)}}

\def \H           {\hat{H}}
\def \T           {\hat{T}}
\def \n           {\ket{\varphi_n}}
\def \ntilde      {\ket{\tilde\varphi_n}}
\def \Emax        {E_{\max}}
\def \tauMT       {\tau_{\mathrm{MT}}}
\def \tauML       {\tau_{\mathrm{ML}}}
\def \tauMLd      {\tau^{\star}_{\mathrm{ML}}}
\def \taubw       {\tau_{\mathrm{bw}}}
\def \tauQSL      {\tau_{\mathrm{QSL}}}
\def \psipsi      {\braket{\psi_0|\psi_t}}
\def \psipsiperp  {\braket{\psi_0|\psi_{t_\perp}}}
\def \psipsiconj  {\braket{\psi_t|\psi_0}}

\begin{document}

\title{Quantum Speed Limit for States with a Bounded Energy Spectrum}

\author{Gal Ness\,\orcidlink{0000-0003-3598-134X}}
\affiliation{Physics Department and Solid State Institute, Technion --- Israel Institute of Technology, Haifa 32000, Israel}
\author{Andrea Alberti\,\orcidlink{0000-0002-1698-3895}}
\affiliation{Fakultät für Physik, Ludwig-Maximilians-Universität München, 80799 München, Germany}
\affiliation{Max-Planck-Institut für Quantenoptik, 85748 Garching, Germany}
\affiliation{Munich Center for Quantum Science and Technology, 80799 München, Germany}
\author{Yoav Sagi\,\orcidlink{0000-0002-3897-1393}}
\email[Electronic address: ]{yoavsagi@technion.ac.il}
\affiliation{Physics Department and Solid State Institute, Technion --- Israel Institute of Technology, Haifa 32000, Israel}

\date{\today}

\begin{abstract}
Quantum speed limits set the maximal pace of state evolution.
Two well-known limits exist for a unitary time-independent Hamiltonian: the Mandelstam--Tamm and Margolus--Levitin bounds.
The former restricts the rate according to the state energy uncertainty, while the latter depends on the mean energy relative to the ground state.
Here we report on an additional bound that exists for states with a bounded energy spectrum.
This bound is dual to the Margolus--Levitin one in the sense that it depends on the difference between the state's mean energy and the energy of the highest occupied eigenstate.
Each of the three bounds can become the most restrictive one, depending on the spread and mean of the energy, forming three dynamical regimes which are accessible in a multi-level system.
The new bound is relevant for quantum information applications, since in most of them, information is stored and manipulated in a Hilbert space with a bounded energy spectrum.
\end{abstract}

\maketitle

Quantum speed limits restrict the ultimate performance of any quantum device by setting a bound to the maximal rate of state evolution \cite{Deffner2017}.
The first of those limits was posed by Mandelstam and Tamm (MT), who realized that the time-energy ``uncertainty relation'' $\Delta E\Delta t\geq\hbar/2$ is not a statement about simultaneous measurements of observables,
but rather about the intrinsic timescale of unitary evolution in quantum mechanics \cite{Mandelstam1945}.
This relation restricts the minimal time for a unitary system to propagate between two states, depending on its energy spread $\Delta E\equiv({\braket{\H^2}-\braket{\H}{}^{\hspace{-2pt}2}}){}^{1/2}$.
An additional independent limit was found by Margolus and Levitin (ML) \cite{Margolus1998}.
The ML bound depends on the mean energy, $E\equiv\braket{\H}$, measured relative to the energy of the lowest occupied eigenstate (e.g., the ground state).
Levitin and Toffoli showed that the unified bound is tight \cite{Levitin2009}.
That is, the time required for arriving at an orthogonal state $t_{\perp}$ is bounded from below by 
\begin{equation}\label{eq:tau_QSL}
t_{\perp}\ge\tauQSL=\max\left\{\tauMT,\tauML\right\}\,.
\end{equation}
Here, $\tauMT\equiv\pi\hbar/(2\Delta E)$ and $\tauML\equiv\pi\hbar/(2E)$ are the minimal orthogonalization times due to the MT and ML bounds, respectively.

In this letter, we introduce a new bound that holds for any state whose spectral decomposition is bounded from above.
We call it the \emph{dual ML bound} because it is essentially equivalent to the ML bound in time-reversed dynamics.
The dual ML bound is determined by the difference between the upper bound and the mean energy.
It applies to a wide range of scenarios, including any upper-bounded Hamiltonian, states composed of a finite number of eigenstates, band-insulator states, and states that populate a UV-cropped set of levels within an unbounded spectrum.
After proving the bound, we discuss the different dynamical regimes in systems with two and three levels.
In particular, we show under what conditions the dual bound dominates the dynamics.
Finally, we discuss the requirements for the experimental probe of the new bound and the three dynamical regimes.

Before proving the new limit, we first briefly recall the original derivation of the ML bound \cite{Margolus1998,Deffner2017}.
We consider the evolution under a time-independent Hamiltonian $\H$.
The time-evolved state at time $t$ can be written as ${\ket{\psi_{t}}=\sum_{n=0}^{\infty}c_{n}e^{-iE_{n}t/\hbar}\n}$, with $c_n\in\mathbb{C}$ such that ${\sum_{n}\left|c_n\right|^2=1}$, and $\n$ are the eigenstates of $\H$ with eigenenergies $E_n\,:\,\H\n=E_n\n$.
Since energy is defined up to an additive constant, the ground state energy can be chosen such $E_n\ge E_0 = 0\;\forall n$ without loss of generality.

To find a lower bound for the orthogonalization time, we consider the two-time state overlap,
${\psipsi=\sum_{n=0}^{\infty}\left|c_n\right|^2 e^{-iE_n t/\hbar}}$, and invoke the trigonometric inequality $1-\cos x\le \frac{2}{\pi}\left(x+\sin x\right)$, which is valid for any $x\ge0$.
We thus obtain
\begin{align} 
1-\mathrm{Re}\left[\psipsi\right]
& =\sum_{n=0}^{\infty} \left|c_n\right|^2 \left(1-\cos\frac{E_{n}t}{\hbar}\right)\nonumber \\
& \le \sum_{n=0}^{\infty}\left|c_n\right|^2\frac{2}{\pi}\left(\frac{E_{n}t}{\hbar}+\sin\frac{E_{n}t}{\hbar}\right)\nonumber \\
& =\frac{2}{\pi}\left(\frac{Et}{\hbar}-\mathrm{Im}\left[\psipsi\right]\right)\,,
\label{eq:ineq1}
\end{align}
where $E=\sum_{n=0}^{\infty}\left|c_{n}\right|^{2}E_n$ is the energy expectation value.
When the system reaches an orthogonal state, i.e., $\psipsiperp=0$, we have $\mathrm{Re}\left[\psipsiperp\right]=\mathrm{Im}\left[\psipsiperp\right]=0$.
Hence, Eq.\,\eqref{eq:ineq1} reduces to $1\le\frac{2}{\pi}\frac{E}{\hbar}t_{\perp}$, implying $t_{\perp}\ge\tauML$.

We now turn to prove the dual bound based on the previous derivation.
Let us consider a state with an energy spectrum bounded from above, with maximum eigenenergy $\Emax$, such that $E_n \in \left[0,\Emax \right] \;\forall n$.
Relying on the same trigonometric inequality used before, we obtain
\begin{align}
& 1-\mathrm{Re}\bigl[ e^{-i \Emax t/\hbar} \psipsiconj \bigr]\nonumber \\
& \quad=\sum_{n=0}^{\infty}\left|c_n\right|^2\left[1-\cos\frac{\left(\Emax-E_n\right)t}{\hbar}\right]\nonumber \\
& \quad\le\frac{2}{\pi}\left\{\frac{\left(\Emax-E\right)t}{\hbar}-\mathrm{Im}\bigl[e^{-i \Emax t/\hbar} \psipsiconj\bigr]\right\}\,.
\end{align}
Considering the time $t_\perp$, when $\ket{\psi_{t_\perp}}$ is orthogonal to the initial state, this inequality reduces to ${1\leq 2 (\Emax-E)t_{\perp}/ (\pi\hbar)}$.
This yields a new bound on the minimum orthogonalization time,
\begin{equation}
t_{\perp}\ge\tauMLd\equiv\frac{\pi\hbar}{2\left(\Emax-E\right)}\;,\label{eq:tauMLd}
\end{equation}
where $\tauMLd$ is the dual ML orthogonalization time.
Equation~\eqref{eq:tauMLd} is the main result of this work.
With this additional bound, the unified quantum speed limit Eq.\,\eqref{eq:tau_QSL} can be generalized to 
\begin{equation}
t_\perp \ge \tauQSL=\max\left\{\tauMT,\tauML,\tauMLd\right\}\,.\label{eq:tauWithDual}
\end{equation}
Note that the dual ML bound also applies when the spectrum of the Hamiltonian is unbounded, as long as the state $\ket{\psi_t}$ has a bounded spectral decomposition.

The ML bound was generalized to a family of bounds:
${t_\perp \ge \tau_{\mathrm{ML},p} \equiv \pi\hbar/(2^{1/p}E_p)}$, where ${E_p\equiv\braket{ (\H-E_0)^p}{}^{\!\!1/p}}$ is the L$^p$-norm of $\hat{H}-E_0$ \cite{Luo2005,Zielinski2006}.
Applying the time-reversal consideration to the derivation of Ref.~\cite{Zielinski2006} yields a family of generalized dual bounds:
${t_\perp \ge \tau^\star_{\mathrm{ML},p} \equiv \pi\hbar/(2^{1/p}E^\star_{p})}$, with ${E_p^\star \equiv \braket{(\Emax-\H)^p}{}^{\!\!1/p}}$.
Compared to $\tauML$ and $\tauMLd$, the generalized bounds can be tighter, depending on the specific spectral distribution of $\ket{\psi_t}$.
However, $E_p$ is not readily associated with a measurable quantity, unlike $E$ and $\Delta E$ \cite{Ness2021}.
For this reason, the case $p>1$ may be less useful in practice.

In the limit $p\rightarrow \infty$, it follows $E_{\infty} = \Emax-E_0 \equiv E_\mathrm{bw}$, which expresses a measure of the bandwidth.
The corresponding bound, $t\ge\pi\hbar/E_\mathrm{bw}\equiv \taubw$, is aptly called the bandwidth speed limit \cite{Boixo2007,Chenu2017,Margolus2021,Aifer2022}, and it constitutes its own dual ($\tau^\star_{\mathrm{bw}} = \taubw$).
We note that this limit can be readily derived as a corollary of the dual ML bound:
Because the average energy lies in either the lower or upper half of the spectrum, we have the inequality $E_\mathrm{bw} \ge 2\,\min\left\{ E-E_0,\Emax-E \right\}$.
Combining this with Eq.\,\eqref{eq:tauWithDual} gives $t_{\perp}\ge \tauQSL \ge \taubw\equiv\pi\hbar/E_\mathrm{bw}$.
These inequalities also show that the bandwidth limit is less tight than Eq.\,\eqref{eq:tauWithDual}.
This limit has the advantage, however, of not requiring any knowledge about the state apart from its spectral boundaries.

To get a deeper insight into the dual bound, we exploit a time-reversal symmetry of the ML bound.
We consider a generic antiunitary operator $\T$.
Its action on the state $\ket{\psi_t}$ is to map ${t\mapsto-t}$ (time-reversal).
By also shifting the energy reference by $\Emax$, we thus get
\begin{align}
\ket{\tilde \psi_t} &\equiv e^{-i \Emax t/\hbar} \T \ket{\psi_{t}}\nonumber\\
& = \sum_{n=0}^{\infty}c_{n} e^{-i (\Emax-E_n) t/\hbar} \ntilde\,,\label{eq:reversed_dynamics}
\end{align}
where $\ntilde \equiv \hat{T} \n$ is a new basis of states.
Since, by definition, $\ket{\psi_{t_\perp}}$ is orthogonal to $\ket{\psi_0}$, it directly follows that $\ket{\tilde \psi_{t_\perp}}$ is also orthogonal to $\ket{\tilde \psi_{0}}$, and \emph{vice versa}.
Therefore, we obtain an equivalent and alternative picture where the considered times are positive, and the energy spectrum is inverted, $E_n\mapsto\Emax-E_n$.
Accordingly, the upper bound of the spectrum $\Emax$ corresponds to a lower bound on the inverted spectrum, and the dual ML bound can be interpreted as the original one under the application of the time-reversal operator.
Hence, by applying the original ML limit \cite{Margolus1998} to the time-reversed state in Eq.\,\eqref{eq:reversed_dynamics}, we directly obtain Eq.\,\eqref{eq:tauMLd}.
We note that in contrast to the ML bound, the derivation above does not yield a new result in the case of the MT bound because $\Delta E$ is invariant under the application of the time-reversal operator $\T$.

Quantum speed limits can be extended to account for the case where the system does not evolve to an orthogonal state.
To this end, one considers the absolute value of the two-time state overlap, $\left| \psipsi\right|$, which gives a measure of how far the system has evolved from the initial state \footnote{The two-times state overlap, which represents the survival amplitude, is also related to the geodesic distance of the Fubini-Study metric between the initial state and the time-evolved state, via $\arccos \left|\psipsi\right|$.}.
The generalized MT bound was shown by Fleming \cite{Fleming1973} to be of the form
${\arccos \left|\psipsi\right|\le \frac{\pi}{2}t/\tauMT}$.
The ML bound was extended to arbitrary states by Giovannetti \emph{et al.} \cite{Giovannetti2003} using the generalized inequality $1-\cos x\le a x + q\sin x$, which holds for ${x\ge0\,\cap\, q\ge0}$, with $a$ being a function of $q$ implicitly defined through a set of equations.
The extended limit can be written as $\arccos \left|\psipsi\right|\le \frac{\pi}{2}\left(t/\tauML\right)^{1/2}\xi(t/\tauML)$,
where $\xi(x)$ is a function very close to unity. The approximation $\xi(x)\approx 1$ is often made in the literature \footnote{We compute and employ a numerical approximation, ${\xi(x) = 1 - 0.0395 (1-x) + \delta(x)}$, with $0\leq \delta(x)<5\times10^{-4}$ being a small correction that approaches 0 for $x\rightarrow 1$.}.

Like for the original ML bound, we obtain an extension of the dual ML bound to non-orthogonal states by applying the derivation from Giovannetti \emph{et al.}\ \cite{Giovannetti2003} to the time-reversed state $\ket{\tilde \psi_t}$ with the energy spectrum $\tilde E_n$.
Hence, combining the extended dual bound with the previous results, we find that the minimum time to attain a certain overlap $|\psipsi|$ is limited by
\begin{multline}
\arccos\left|\psipsi\right|\le\frac{\pi}{2}\min\left\{ \frac{t}{\tauMT}\,,\right.\\
\left.\xi\left(\frac{t}{\tauML}\right) \sqrt{\frac{t}{\tauML}}\,,\;\xi\left(\frac{t}{\tauMLd}\right) \sqrt{\frac{t}{\tauMLd}}\right\}\,.
\label{eq:QSLwithDual}
\end{multline}

\begin{figure}
\centering
\includegraphics{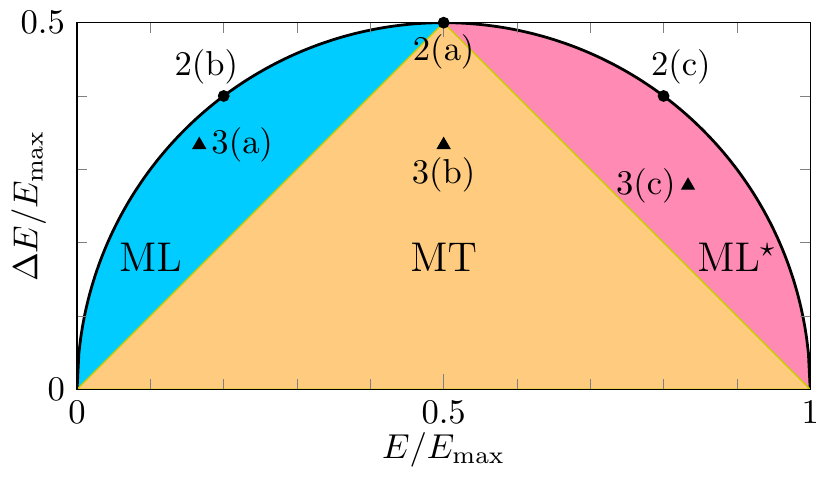}
\caption{\textbf{Dynamical regimes of quantum state evolution}.
The dynamical regimes are displayed on the energy uncertainty ($\Delta E$) vs. mean energy ($E$) graph.
Three regimes are identified depending on which of three terms in Eq.\,\eqref{eq:QSLwithDual} is most constraining:
the yellow area is the MT regime, the blue area is the (original) ML regime, and the red area is the regime associated with new dual ML bound.
The solid black line denotes the maximal energy uncertainty as a function of mean energy when the spectrum is bounded; see Eq.\,\eqref{eq:deltaEmax}.
The circles and triangles are followed by labels referring to the examples presented in Figs.~\ref{fig:qubitEvolutions} and \ref{fig:qutritEvolutions}, respectively.
Note that the point 2(a) is a singular point where all three regimes touch.
\label{fig:QSLregimes}}
\end{figure}

The unified limit in Eq.\,\eqref{eq:QSLwithDual} gives rise to three dynamical regimes, as presented in Fig.~\ref{fig:QSLregimes}.
For states with ${\Delta E<\min\left\{E,\Emax-E\right\}}$, we have ${\tauMT>\max\left\{\tauML,\tauMLd\right\}}$, and the energy uncertainty, $\Delta E$, is the only relevant quantity determining the timescale of the dynamics.
This is the MT regime denoted by the yellow shading.
For states with ${\Delta E>\min\left\{E,\Emax-E\right\}}$, we get two other regimes.
In the ML regime, when $E<\Emax/2$ (blue shading), the overlap is restricted by the MT bound for initial times, and by the ML bound for times longer than a crossover time defined by $\tau_{c}\equiv\xi^2\tauMT^2/\tauML$.
Similarly, in the regime of the dual ML bound, when $E>\Emax/2$, a crossover occurs at $\tau^\star_{c}\equiv\xi^2\tauMT^2/\tauMLd$.
The diagram shows that the dual ML bound defines a limit on quantum state evolution that is complementary to the original ML bound.

The diagram in Fig.~\ref{fig:QSLregimes} is further constrained by Popoviciu's inequality \cite{Popoviciu1935,Aifer2022}, 
\begin{equation}
\Delta E\le\sqrt{E\left(\Emax-E\right)}\;.\label{eq:deltaEmax}
\end{equation}
This additional constraint is depicted by a black line in Fig.~\ref{fig:QSLregimes}, with the forbidden region painted in white.
Note that the inequality in Eq.\,\eqref{eq:deltaEmax} is saturated if and only if the state comprises only two levels (a qubit or an effective qubit in a multi-level system).
This fact implies that a qubit is always in either the ML regime or in the dual ML one.
In order to access the MT regime, in which the evolution is solely limited by the MT bound, one has to consider a state of at least three occupied energy levels.

\begin{figure}
\centering
\includegraphics{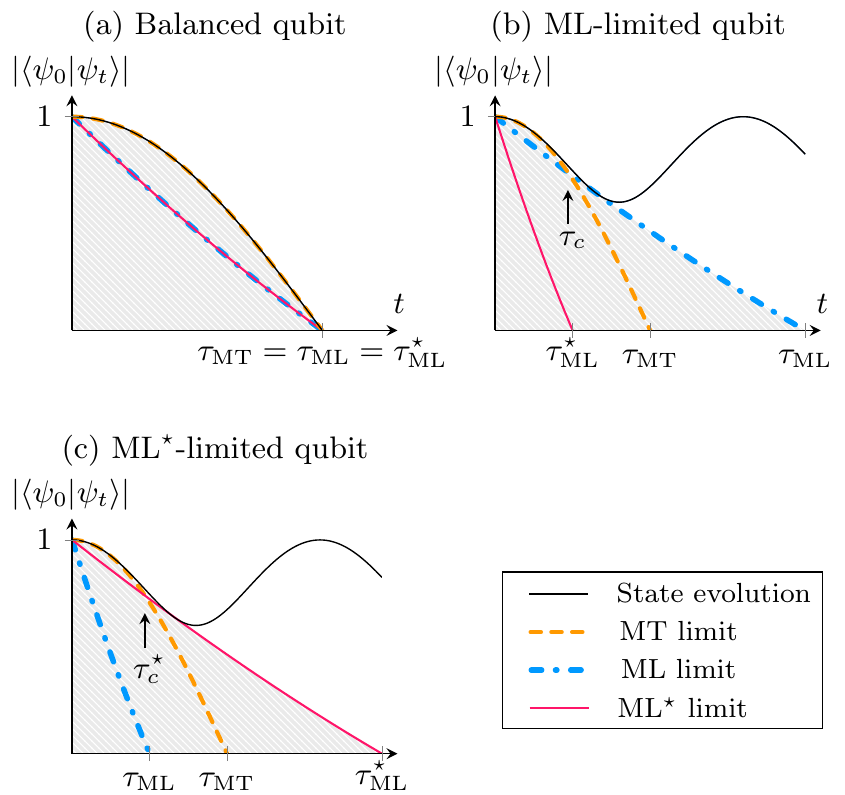}
\caption{\textbf{Qubit system evolution in different dynamical regimes}.
The amplitude of the two-time state overlap and its limits are plotted as a function of time.
The black line represents the state evolution, whereas the yellow, blue, and red lines denote the MT, ML, and dual ML bounds.
The region excluded by the unified bound in Eq.\,\eqref{eq:QSLwithDual} is indicated by the gray area.
(a) The case of $E=\Delta E$ is the only one saturating both bounds.
(b) For a qubit state with $E<\Delta E$, the evolution is initially limited by the MT bound and at later times by the ML one.
(c) A state with $E>\Emax-\Delta E$ has its evolution restricted by the dual ML bound for times $t>\tau_c^\star$.
\label{fig:qubitEvolutions}}
\end{figure}

To illustrate the implications of the new bound in Eq.\,\eqref{eq:QSLwithDual},
we consider first the dynamics of a single qubit.
The state is given by ${\ket{\psi}=c_0\ket{0}+c_1\ket{1}}$, with ${|c_0|^2+|c_1|^2=1}$, and the Hamiltonian is $\H=\Emax\ket{1}\!\bra{1}$.
Three different scenarios are presented in Fig.~\figref{fig:qubitEvolutions}{a--c}, with the black curves showing the overlap's amplitude evolution, which is bounded by the MT (yellow), ML (blue), and dual ML (red) limits.
The energy moments of these scenarios are marked by solid circles in Fig.~\ref{fig:QSLregimes}.

Figure~\figref{fig:qubitEvolutions}{a} shows the special case of a balanced qubit ($c_0=c_1$) for which $E=\Delta E=\frac{1}{2}\Emax$.
In this case, the MT bound coincides with the two-time state overlap for all times $0\le t\le \tauMT$ \cite{Levitin2009}.
The balanced qubit is the only case for which all three limits coincide at a certain time, which is also the orthogonalization time $t_{\perp}=\tauQSL$.
In contrast, when $c_0=2c_1$, we have $E<\Delta E$, since
$E=\frac{1}{2}\Delta E=\frac{1}{5}\Emax$.
In this scenario, a crossover to the ML bound occurs for $t>\tau_c$, as shown in Fig.~\figref{fig:qubitEvolutions}{b}.
If instead we consider the dual qubit, with ${2c_0=c_1}$,
the complementary condition is realized, ${E>\Emax-\Delta E}$, since ${E=2\Delta E=\frac{4}{5}\Emax}$.
This scenario is presented in Fig.~\figref{fig:qubitEvolutions}{c}, revealing a similar dynamics to that shown in Fig.~\figref{fig:qubitEvolutions}{b}.
For the dual qubit, however, the quantum state evolution is constrained by the dual ML bound at times larger than $\tau^\star_{c}$, and never by the original ML bound.
Note that the two-time state overlap $\left|\psipsi\right|$ saturates the original ML bound [panel (b)] for some time $\tau_c\le t \le \tauML$; same for the dual ML bound [panel (c)] for a certain time $\tau_c^\star\le t \le \tauMLd$ \cite{Giovannetti2003}.

As previously noted, a multi-level system is necessary to appreciate the MT regime.
Accordingly, we now consider a three-level Hamiltonian (qutrit), ${\H=\eta\Emax\ket{1}\!\bra{1}+\Emax\ket{2}\!\bra{2}}$ with $\eta\in\left(0,1\right)$.
In this system, a general state is given by ${\ket{\psi}=c_0\ket{0}+c_1\ket{1}+c_2\ket{2}}$, where the coefficients can be related to the energy moments $E$ and $\Delta E$:
\begin{equation}
\begin{cases}
\left|c_0\right|^2=1-\left|c_1\right|^2-\left|c_2\right|^2\,,\\
\left|c_1\right|^2=\frac{1}{\left(1-\eta\right)\eta}\left[\left(1-\frac{E}{\Emax}\right)\frac{E}{\Emax}-\frac{\Delta E^{2}}{\Emax^{2}}\right]\,,\\
\left|c_2\right|^2=\frac{1}{1-\eta}\left[\left(\frac{E}{\Emax}-\eta\right)\frac{E}{\Emax}+\frac{\Delta E^{2}}{\Emax^{2}}\right]\,.
\end{cases}
\end{equation}

\begin{figure}
\centering
\includegraphics{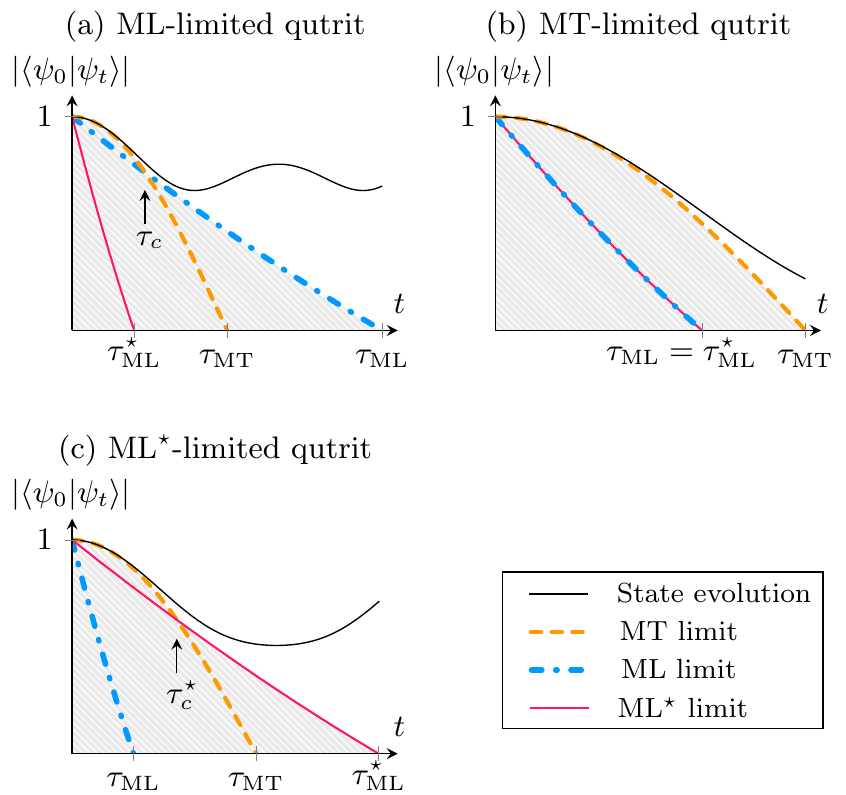}
\caption{\textbf{Qutrit system evolution in different dynamical regimes}.
(a) For a qutrit state with $E<\Delta E<(\Emax-E)$, the evolution is initially limited by the MT bound and at later times by the ML one.
(b) A state with $\Delta E<E$ is exclusively restricted by the MT bound.
(c) For a state with $(\Emax-E)<\Delta E<E$, the dual ML bound is the one constraining the dynamics for times $t>\tau_c^\star$.
\label{fig:qutritEvolutions}}
\end{figure}

Three paradigmatic cases are presented in Fig.~\ref{fig:qutritEvolutions}, with the same convention of colors and symbols used in Fig.~\ref{fig:qubitEvolutions}.
For simplicity, we set $\eta=\nicefrac{1}{2}$ in all examples.
Considering the state with $E=\frac{1}{2}\Delta E=\frac{1}{6}\Emax$, the evolution is initially bounded by the MT bound and later by the ML one, as depicted in Fig.~\figref{fig:qutritEvolutions}{a}.
For comparison, we show in Fig.~\figref{fig:qutritEvolutions}{b} the case of a qutrit state with $E=\frac{3}{2}\Delta E=\frac{1}{2}\Emax$.
In this case, the MT limit is the only bound to participate in the dynamics.
Instead, for a qutrit state with $E=3\Delta E=\frac{5}{6}\Emax$, the quantum state evolution is constrained by the dual ML bound at later times, when $t>\tau_c^\star$, as presented in Fig.~\figref{fig:qutritEvolutions}{c}.
These three qutrit scenarios are marked by solid triangles in Fig.~\ref{fig:QSLregimes}.
The qutrit examples show that all the dynamical regimes depicted in Fig.~\ref{fig:QSLregimes} can occur in real systems.

In a multiple-level system, the situation falls back to the case of a single qubit when the quantum state $\ket{\psi_t}$ can be written as a sum of two eigenstates.
In particular, when the superposition of the two eigenstates is balanced, all three terms in the unified limit in Eq.\,\eqref{eq:QSLwithDual} are saturated at $t=t_{\perp}$.
A balanced superposition of two eigenstates can be realized in a system of multiple noninteracting qubits when either a single qubit is allowed to evolve while the rest are frozen, or when the system is prepared in a maximally entangled state of the GHZ form \cite{Giovannetti2003a}.
Alternatively, a balanced superposition saturating the limit in Eq.\,\eqref{eq:QSLwithDual} is realized when the qubits can interact through a nonlocal Hamiltonian \cite{Bukov2019}.

Finally, we discuss the possibility of observing the dual ML bound in an experiment.
As shown above, the dual ML bound can be explored with a simple qubit system.
However, to probe \emph{bona fide} the MT regime, it is necessary to control at least three energy levels.
In particular, it must be possible to vary their populations in order to tune the ratios between the three orthogonalization times.
Moreover, the state has to feature an energy cutoff, where only levels with energy lower than $\Emax$ are populated.
For the detection, methods should be available to measure the mean energy, the energy uncertainty and, most importantly, the two-time state overlap.

Recently, the experimental measurement of the ML and MT bounds was reported with a spatially excited state of a single atom in a one-dimensional optical lattice \cite{Ness2021}.
The method leverages an atomic interferometer, where in one branch the atom is prepared by a rapid quench of the Hamiltonian into an excited state, thus populating many motional eigenstates, whereas in the second branch the atom remains stationary in a motional eigenstate and serves as a reference for the initial state $\ket{\psi_0}$.
The interference between the two branches provides a measurement of the two-time state overlap $\psipsi$.
In this system, the finite trap depth induces an effective energy cutoff, where the excitation of unbounded states causes the atoms to escape from the trap.
Thus, one could effectively avoid the excitation of unbound states by renormalizing the contrast of the Ramsey interferometer to account for the loss of escaped atoms.

Several other systems are also candidates for testing the unified bound in Eq.\,\eqref{eq:QSLwithDual} with three controlled levels relying, among others, on single photons \cite{Langford2004,Luo2019}, trapped ions \cite{Klimov2003}, trapped atoms \cite{Belmechri2013}, and artificial atoms in superconducting quantum circuits \cite{Bianchetti2010,Vepsalainen2019,CerveraLierta2022}.

To summarize, we have shown that the rate of evolution of an energy bounded state under the effect of a time-independent Hamiltonian is constrained by a bound dual to the ML one. Accordingly, the minimal time to go from an initial state to an orthogonal time-evolved state is restricted by three factors: the energy uncertainty (MT bound), the mean energy difference from the lowest occupied eigenenergy (original ML bound), and its difference from highest populated eigenenergy (dual ML bound). When the energetic difference of the state from the closest end of the populated spectrum is smaller than the energy uncertainty, it is in the ML or dual ML regimes, which feature a two-staged evolution: in the first, the quantum evolution is constrained by the MT bound and in the second by either one of the ML bounds.

At the foundation of quantum speed limits, there is the trade-off between the amount of information known about the system and how tight the bound is.
On one extreme, there is the bandwidth limit \cite{Boixo2007,Chenu2017,Margolus2021,Aifer2022}, which utilizes minimal knowledge about the state but produces a loose limit.
Conversely, generalized bounds \cite{Luo2005,Zielinski2006,Chau2010} capture the dynamics more closely, but require substantial information about the spectral decomposition of the state.
The dual ML, combined with the original limits, form a set of speed limits that offers the tightest unified bound that can be constructed from the first two energy moments of the Hamiltonian --- two quantities readily accessible from experiments \cite{Ness2021}.

\begin{acknowledgments}
We thank Manolo R.~Lam, Joseph E.~Avron, Dieter Meschede, Norman H.~Margolus, Adolfo del Campo, Sebastian Deffner and Steve Campbell for helpful discussions.
This research was supported by the Reinhard Frank Foundation in collaboration with the German Technion Society, the Israel Science Foundation (ISF), grant No.~3491/21, and by the Pazy Research Foundation.
\end{acknowledgments}

\end{document}